\documentstyle[times,namedreferences]{spackap}
\begin{opening}
\title{AB Dor: A Single Star with RSCVn Like Activity in X-ray Band}
\author{M. \surname{Alev}}
\author{M. A. \surname{Alpar}}
\author{A. \surname{Esendemir}}
\author{\"U. \surname{KIzIlo\u{g}lu}}
\institute{Physics Department, Middle East Technical University, Ankara 06531, Turkey }
\date{}
\end{opening}
\begin{document}

\begin{abstract}
Using the archival ROSAT PSPC observations, AB Dor is found to be 
variable in X-rays. The periodic variations are consistent with 
previously reported rotational period of 0$^d$.514. The average spectrum
of AB Dor is best represented with two-temperature 
Raymond-Smith model with kT values of 0.19$\pm$0.07 and 
1.17$\pm$0.02 keV. The quiescent luminosity of the system is found to be  
4.36$\pm$0.6$\times$10$^{30}$ ergs s$^{-1}$. A flare with 
a rise time of $\sim$350 seconds is detected during which X-ray luminosity
rises from 5.8$\pm$1.6$\times$10$^{30}$ to 15.8$\pm$4.9$\times$10$^{30}$
ergs s$^{-1}$. We conclude that AB Dor is very similar to the active
components of RS CVn binaries and other active classes. In view of the wide 
separation from the binary companion Rst 137B, this activity must be 
intrinsic to the active star.
\end{abstract}

\def\exo{{\sl EXOSAT }}
\def\ein{{\sl EINSTEIN }}
\def\uhu{{\sl UHURU }}
\def\ro{{\sl ROSAT }}
\def\ergsec{\hbox{erg s$^{-1}$ }}
\def\ergcm{\hbox{erg cm$^{-2}$ s$^{-1}$ }}
\def\la{\raise.5ex\hbox{$<$}\kern-.8em\lower 1mm\hbox{$\sim$}}
\def\ma{\raise.5ex\hbox{$>$}\kern-.8em\lower 1mm\hbox{$\sim$}}
\def\ea{\it et al. \rm}
\def\am{$^{\prime}$\ }
\def\as{$^{\prime\prime}$\ }
\def\eg{{\sl EGRET }}
\section{Introduction}
 AB Dor is a very active K type star in the southern hemisphere. From the
proper motion and radial velocity measurements, its distance is estimated to
be 27$\pm$7 pc (Collier Cameron et al., 1988) so that it is a foreground star in LMC region. The star
was classified as an RS CVn type binary by Pakull(1981) because of its
photometric variability.  AB Dor is a member of a wide visual binary system
with its companion Rst 137B, a dMe type 13.6 magnitude star. Both stars have common proper motions and radial velocities. Thus they are probably the members of
the Local Association  Pleiades moving group (Vilhu et al., 1993). 
The wide separation of the binary makes AB Dor an interesting candidate
for comparison with active close binaries, as the activity must arise
from the dynamics of the single star.
AB Dor is a rapidly rotating star. 
From the broadening of absorption
lines (Collier, 1982), its equatorial velocity is found to be 70$\pm$5 km s$^{-1}$.
 Because of its relative proximity and
brightness (V=6.8 mag), AB Dor has been observed photometrically by various
observers (Pakull, 1981; Rucinski, 1983; Innis et al., 1988; 
Banks et al., 1991; Jetsu et al., 1990; Rucinski et al., 1995). 
In a photometric history
study by Innis et al.(1988) which shows compiled light curves between
1978 and 1987, the rapid changes in the shape and range of light curves
have been noted. 
Light curves of AB Dor show significant changes even in  time intervals 
as short as only a few rotational
periods of the star thus making it difficult to combine the photometric data
taken a few months apart to obtain a consistent light curve. These changes are
thought to originate from the rapid rearrangements of spotted areas (Jetsu et
al., 1990).

 During an X-ray survey of the Large Magellanic Cloud with the imaging X-ray
telescope of the Einstein Observatory, AB Dor was identified to be one of the
brightest X-ray point sources in the foreground region 
 in 0.15-4.5 KeV band (Long et al., 1981). A
subsequent three hour observation  has revealed clear 
indications of variability. A flare of two-hour duration,  with nearly
five-fold increase in the x-ray flare count rate, has been detected
with the Einstein Observatory (Pakull, 1981). AB Dor was also observed by
low and medium energy detectors of EXOSAT spacecraft in late 1984 and early
1986. In each observation covering nearly one rotational period of the
star, X-ray flares with long rise times ($\sim 6000 s$) were detected (Collier
Cameron et al., 1988). From the long rise times of the observed X-ray flares, it
was suggested that the flares occurred in loop structures of several stellar
radii. 
 In 1990, AB Dor was observed by Large Area Counter Instrument of the GINGA
satellite (Vilhu et al., 1993). During the period covering nearly five stellar
rotations of the star, four flares were detected. Mean flare energies were 
found to be $(1-2)\times 10^{34}$ ergs  
with peak luminosities of $(4-6)\times
10^{30}$ ergs s$^{-1}$. The flare spectra were seen to be best-fitted by thermal
Bremsstrahlung model with kT 4.6 keV. Due to the low sensitivity of the instrument
no clear modulation at the 0$^d$.514 rotation period was reported in the 
X-ray light curve.
 Molonglo Observatory Synthesis Telescope (MOST) observations of AB Dor   
since 1985 have revealed a quiescent level of radio emission with occasional 
flare-like increases in the 843 MHz band (Beasley et al., 1993). 
AB Dor is thoroughly investigated in a multi-wavelength campaign in 1994. A summary of ROSAT observations of AB Dor is presented in the International Conference on X-ray Astronomy and Astrophysics held in Wuerzburg (Kuerster et al, 1995)

In this work we examine the  X-ray modulation and flaring
structure of AB Dor with archival ROSAT Position Sensitive Proportional 
Counter(PSPC) data.  In section 2 observations 
are summarized.  In Section 3 X-ray spectra are discussed, modulation at 
the 0$^d$.514  period is examined and folded light curves are given. 
Behavior of  X-ray flares is also studied in 
this Section.  A general discussion of X-ray behavior of the star is 
given in section 4.
\section{Observations}
AB Dor was observed with the PSPC at the focus of the X-ray 
telescope of ROSAT.The PSPC is a gas filled proportional counter sensitive 
over the
energy range 0.1-2.4 keV with an energy resolution $\Delta$E/E$\sim$0.43
at 0.93 keV.
Detailed descriptions of the satellite,
X-ray mirrors, and detectors can be obtained in Tr{\"u}mper (1983)
and Pfeffermann et al. (1986).
The X-ray observations reported here were obtained during a time span of 2.7 years between
Oct 28th, 1991 (JD 2448557.5)  and Nov 3rd, 1993 (JD 2449294.5)
with a total effective exposure time of 76210 sec. The journal of the 
observations is given  in Table I. 
\begin{table}
\caption{The journal of ROSAT observations for AB Dor.}
\begin{tabular}{rcr}
\hline\noalign{\smallskip}
 Obs.& Obs. Date  & Eff. Exp.  \\
     &   (JD)     & Time(s) \\
\hline
1  & 2448322.376414002 - 2448322.387212612 &  933  \\
2  & 2448382.727635410 - 2448382.746524297 & 1632  \\
3  & 2448558.053777050 - 2448564.042676994 & 45995  \\
4  & 2448832.757550988 - 2448832.831312557 & 1263  \\
5  & 2448862.506680543 - 2448862.522976838 & 1408  \\
6  & 2448893.285972074 - 2448894.235462740 & 3430  \\
7  & 2448923.594187312 - 2448923.605807682 & 1004  \\
8  & 2448954.529682676 - 2448954.541407212 &  620  \\
9  & 2448954.595226652 - 2448954.610284522 & 1301  \\
10 & 2448984.068661906 - 2448984.211485968 & 2122  \\
11 & 2449014.737230558 - 2449014.753341668 & 1392  \\
12 & 2449044.272980702 - 2449044.284716812 & 1014  \\
13 & 2449075.137226107 - 2449075.211253880 & 1369  \\
14 & 2449104.733902202 - 2449104.749909145 & 1383  \\
15 & 2449135.692557362 - 2449135.925948549 & 1891  \\
16 & 2449165.298955668 - 2449165.460830656 & 2122  \\
17 & 2449196.169694188 - 2449196.184011316 & 1237  \\
18 & 2449225.694865734 - 2449225.741682861 & 989  \\
19 & 2449256.684203886 - 2449256.734944623 & 1466  \\
20 & 2449286.306991192 - 2449286.608171726 & 3639  \\
\hline
\end{tabular}
\end{table}
%
The third observation in the Table I was 
obtained during an  observation of the Large Magellanic Cloud and covers 
$\sim$ 6 days with an effective exposure time of 45995 sec. Observation 
14 in the list is a flare observation with a clear time history and observation
8 is another possible flare but its time history was not followed. 
  The analysis of the ROSAT archival data has been performed with the EXSAS
  package (Zimmermann et al., 1993).
  The AB Dor source counts were extracted from a circle of radius
  $5' $ which is expected to include $99\%$ of the photons from the
source, according to the point spread function of the PSPC.
  The background was determined from a source free annular ring of radius
  $10'$. The mean background
  subtracted, vignetting and deadtime corrected count rate for the whole
  observation was 7.5$\pm$1.5 counts s$^{-1}$. The total light curve of
the source is given in Figure 1.
\begin{figure}
\vspace{8.5cm}
\includegraphics{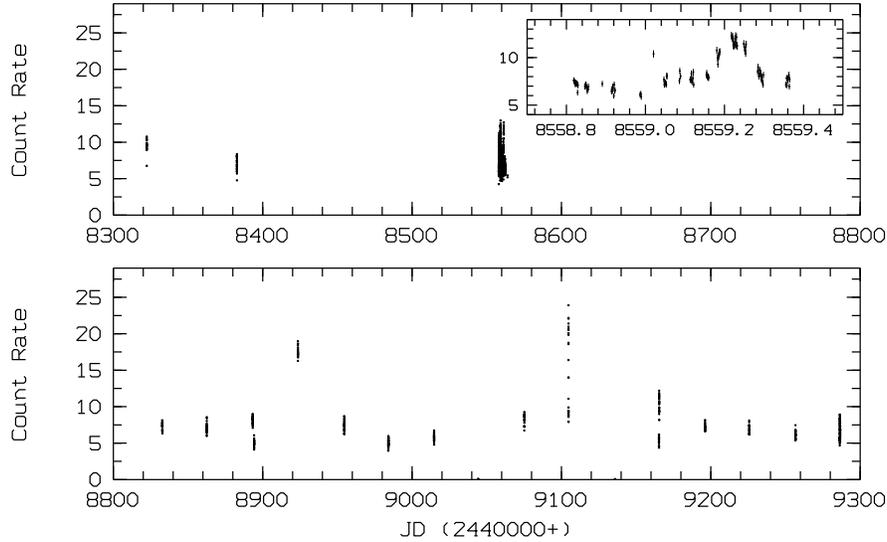}
\caption{The light curve of AB Dor. The observation span is  about 970
days and the compressed region around JD 2448560 is the continuous observation
of AB Dor. The inset shows a portion of compressed part of the light curve. Feature around JD 2449105 is the flare event with clear time 
history.
            }
\label{Fig-1}
\end{figure}
\section{Results}
\subsection{X-Ray Spectra}
The large number of counts obtained from AB Dor allow spectral fits for
various spectral models. 
\begin{figure}
\vspace{11.5cm}
\includegraphics{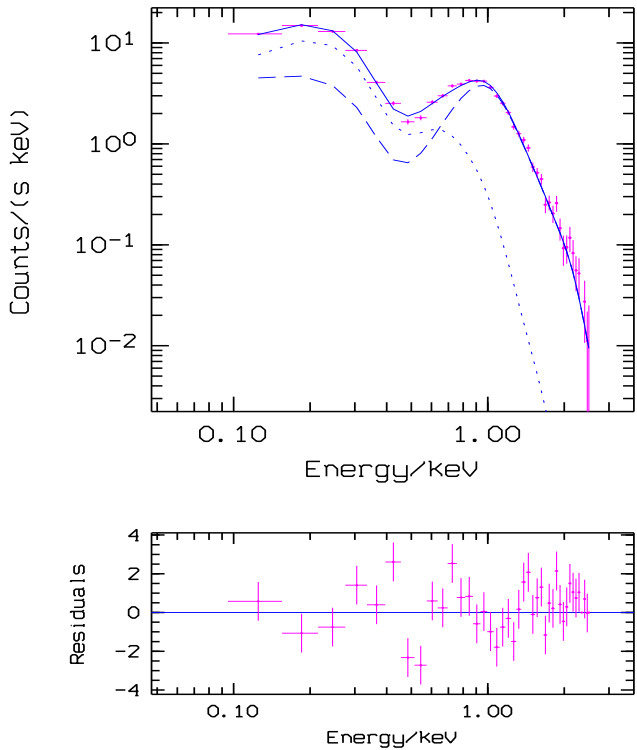}
\caption{The upper panel shows the spectrum of AB Dor as observed with
the PSPC on board ROSAT, together with a two-temperature
 Raymond-Smith spectra  at temperatures kT$_1$=0.19 and
kT$_2$=1.17 keV.
Input spectra is compiled from quiescent regions of the observation 3
of Table 1. The total integration time is $\sim$20000 sec.
($\chi ^{2}_{\nu}=1.78 $ for 34 degrees of freedom). The fitted spectra
is shown as a solid line. The dashed lines are for the two Raymond-Smith
models. The lower panel shows the residuals of the fit.}
\end{figure}

We mainly concentrated on spectral models for thermal plasma:
Raymond-Smith thin plasma, or RS model (Raymond \& Smith, 1977)
and Mewe-Kaastra, or "meka" model (Mewe, Gronenschild, \& van den Oord ,1985;
Kaastra, 1992). Spectral fits carried out by using EXSAS and XSPEC packages have yielded similar results. Single component models failed to represent the observed 
spectra; hence we tried two-temperature RS and "meka"models . The input 
spectra is compiled from the long observation 3 in  Table I. Count rate of the 
source was seen to be highly variable. For this
reason, individual spectral fits of the intervals with high count rate and
low count rate were obtained separately. In this section we discuss the 
spectra for the low count rate (quiescent) intervals.  Total integration time for the
compiled quiescent spectra is $\sim$20000 sec. Although residuals due to some lines in the fitted spectrum can be reduced by using RS variable abundance mode, RS quick mode is used in the analysis because of low energy resolution of ROSAT PSPC. The two-temperature "meka" does not give acceptable $\chi^{2}$ values.
The results of spectral fits are shown in Table II.
\begin{table}
\caption{Spectral Fit Parameters for the observation.}
\begin{tabular}{lcccr}
\hline
Model&N$_{H}$ & (A$_1$/A$_2$)$^{a}$&kT$_1$/kT$_2$&$(\chi^{2}_{\nu})^{b}$ \\
     &$10^{19}$cm$^{-2}$&  &KeV&  \\
\noalign{\smallskip}
\hline
"meka"&1.33$\pm$0.09 &7.09$\pm$0.12$\times10^{-3}$ &0.17$\pm0.05$ & 7.90 \\
    &        &1.97$\pm$0.27$\times10^{-2}$ & 0.87$\pm$0.08   &        \\  
RS & 2.30$\pm$0.69 &6.83$\pm$0.53$\times10^{-3}$ &0.19$\pm$0.07 & 1.78 \\
    &        &1.47$\pm$0.13$\times10^{-2}$ & 1.17$\pm$0.02   &        \\
\noalign{\smallskip}
\hline
\end{tabular}

$^{a}$~~A$_1$ and A$_2$ are the normalization coefficients, 
10$^{-14}$EM/(4$\pi$D$^2$)\\
where D is the source distance and EM is the emission measure,\\
in units of cm$^{-5}$.\\ 
$^{b}$~~Values of reduced $\chi ^{2} _{\nu} $ for 34 degrees of freedom.\\
\end{table}
Figure 2 shows the best fit RS model 
to a typical low-count rate (quiescent) section. 
The time averaged
quiescent luminosity is found to be 4.36$\pm0.6\times10^{30}$ ergs s$^{-1}$. 
\subsection{Rotational Modulation of X-ray Count Rates}
In order to examine whether the X-ray light curve exhibits a 
periodic behavior similar to the optical light curves, the X-ray light curve is
folded by the optical rotational period of 0$^d$.51479  using the
ephemeris given by Innis et al., 1988.  
The longest continuous observation of the source is the third
observation in Table I, between JD 2448558 and JD 2448562, 
covering nearly 8 stellar rotations and hence dominates the whole data. 
Folding procedures are  applied to this continuous part 
of the data and to the rest separately in order to delineate the 
effects of phase drifts. Variability in count rate with rotational phase
is clearly seen in the continuous part of the observation 
in the upper panel of Figure 3. 
\begin{figure}
\vspace{8cm}
\includegraphics{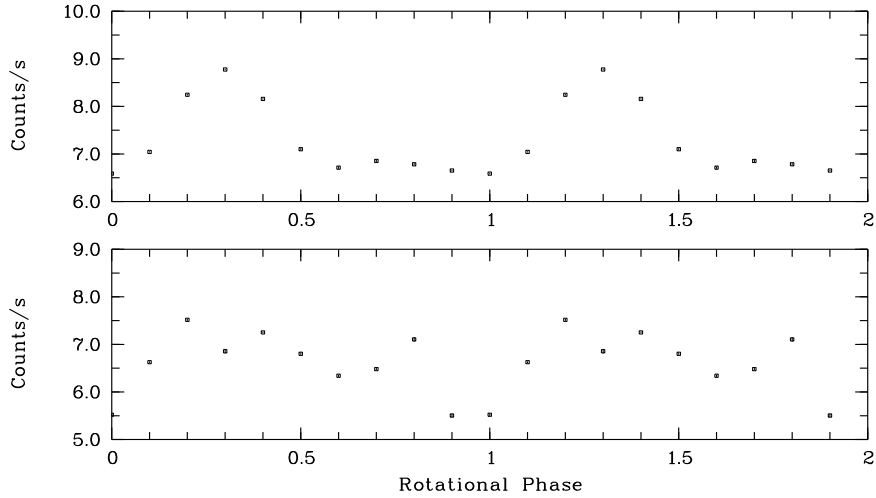}
\caption{Upper panel shows rotational light curve of AB Dor obtained from 
the longest continuous observation of the source between 
JD 2448558.018 and JD 2448562.053 covering nearly 8 rotational periods
folded with the period and the epoch given by Innis et al. (1988). 
Lower panel shows fold of all observation span ($\sim$2.7 years) 
excluding continuous observation and the flare observation 14 of Table I.}
\end{figure}
The X-ray count rate is modulated at about  
23 $\%$ level around the mean count rate of $\sim$7.5 counts s$^{-1}$. The
count rate changes smoothly between the phases 0.0 and 0.5, and
after phase 0.5 the count rate remains almost constant within error limits.
Due to the large number of photons from the source the errors on 
the phase bins are very small. On the other hand, a similar behavior does not 
exist in the rest of the data which is unevenly sampled covering 
$\sim$2.7 years (Figure 3, lower panel). 
A subtler variability  seen might be the result of the fact that 
although the total time span of the observation is as long as 2.7 years, 
the effective exposure time for this part of the data is only $\sim$30000 
seconds  distributed in many windows in the 2.7 years, 
smearing out the basic modulation at 0$^d$.514.
\subsection{X-ray flares}
In the light curve covering nearly a time span of 2.7 years, a few 
probable X-ray flares were detected. The time evolution of the flare can be 
traced clearly in only one flare. We concentrate
on this single flare whose rise time is approximately 350 seconds. A steep
rise in the flux is followed by a rather gentle, slow decline 
(Figure 4 upper panel).  
\begin{figure}
\vspace{9cm}
\includegraphics{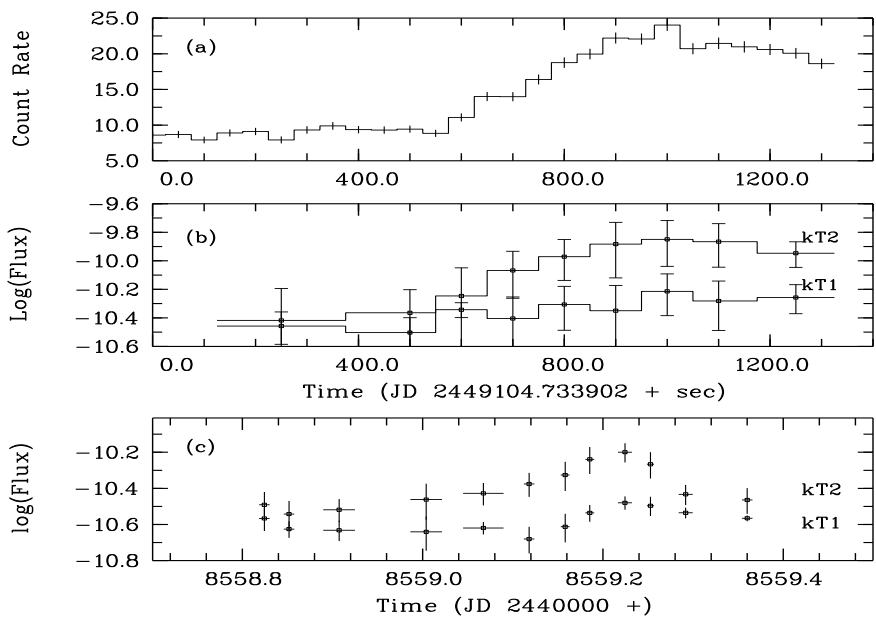}
\caption{Upper panel shows development of flare in observation 14 of
Table 1 and in the middle panel the time history of two different photon
fluxes with energies kT$_1$ and kT$_2$ during the flare development are seen. Flux is the result of the
best fitted two component Raymond-Smith model.The lower panel shows the change of photon fluxes with energies kT$_1$ and kT$_2$ in a high count rate region shown in the inset of Figure 1. }
\end{figure}
The decline  could not be fully observed,
but continued for at least 300 seconds, till the end of the exposure. In
the flare the count rate increases almost three-fold (from $\sim$ 9 to 24
counts s$^{-1}$). The phase of the flare is about 0.06 following the ephemeris 
given by Innis et al. (1988).
To examine the changes in spectra 
during the flare, the ratio of flare to quiescent photons is computed as a function of time.
 A quiescent spectrum
is compiled from the first 500 seconds of the observation prior to the flare. 
It is seen that the counts in channels at larger energy
(channel numbers larger than 100) increase during the
flare  by a factor of $\sim$4 and an increase of $\sim$2 fold
on the lower channels is also evident (Figure 5). 
\begin{figure}
\vspace{9cm}
\includegraphics{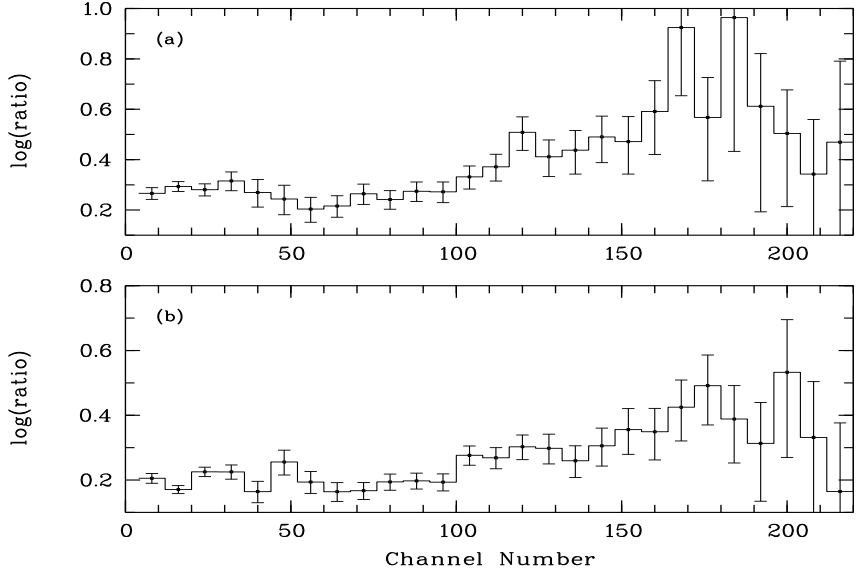}
\caption{The ratio of the flare photon spectra to the quiescent time 
photon spectra.  The quiescent photon spectra compiled from the initial 
500 seconds prior to the flare observation where the count rate is 
Almost constant(upper panel).A similar ratio for a characteristic high count rate region in the inset of Figure 1(lower panel). The quiescent photon spectra was compiled from the region between JD 2448559.20-2448559.28}
\end{figure}
The flare is best-fitted by a 
two-component Raymond-Smith model. 
The time evolutions of the two fit fluxes are given in Figure 4 middle panel. 
The low energy component (changes between kT=0.16-0.24 keV)  can be seen to 
increase very slowly whereas the  
increase in high energy component (kT=1.25-2.34 keV) is much more pronounced. 
No decline can be observed for either component within error limits 
 after 
the peak of the flare is reached.

A sample segment from the light curve of observation 3 showing the transition from low to high count rate is given in the inset of Figure 1.
The section covers a time span of $\sim$50000 seconds, from JD 2448558.8 to
JD 2448559.4 and the maximum count rate coincides with phase 0.36 according
to the ephemeris given by Innis et al. (1988). The rise time from 
low count rate to the
peak is approximately 15000 seconds and a similar decline time is also
observed. The time structure of the flux obtained by the two-component
Raymond-Smith model is shown in the lower panel of Figure 4. 
Although the 
change in the count rate is about two fold as it is seen from Figure 1, the timestructure of the components with energies of kT$_1$ and kT$_2$ 
show similarities with the flare event discussed above. 
The high energy component in this case shows a constant slow increase well before
the increase in the count rate whereas the low energy component follows the
count rate on rise. After the maximum count rate is reached, the high energy 
component follows the count rate on decline, whereas the low energy component
declines rather slowly.
\section{Discussion}
AB Dor is a wide visual binary with an active dMe type dwarf secondary Rst
137B (Vilhu \& Linsky, 1987; Vilhu et al., 1989). 
The angular separation between companion stars is quite large 
(10 arcsec. at a distance of nearly 27 parsecs) and the orbital velocity 
changes of AB Dor is limited to $\pm$2 km s$^{-1}$ (Vilhu et al., 1993).
   Although angular resolution of ROSAT PSPC is better than 
EXOSAT and GINGA, and
both AB Dor and Rst 137B are in the field of view, it is still impossible 
to resolve the companions with PSPC, during the observations. The expected 
X-ray flare peak luminosities from dMe stars is around 
10$^{28-30}$ erg s$^{-1}$ (van den Oord et al., 1988), substantially less
than  the quiescent X-ray luminosity, 4.36$\pm0.6\times$10$^{30}$ ergs s$^{-1}$
mentioned above. Therefore, contamination
due to Rst 137B likely to be  extremely small and its effect can be  neglected.
   The average spectrum of AB Dor, compiled from the quiescent (or the low 
count rate) regions, reveals two distinct regions with temperatures
kT$_1$=0.19 keV and kT$_2$=1.17 keV. 
Column density in 0.1-2.4 keV band is
calculated as N$_H$=2.30$\pm0.79\times10^{19}$ cm$^{-2}$. This value is 
higher than 10$^{18}$ cm$^{-2}$ which is an assumed value used in 
spectral fits to EXOSAT data (Collier Cameron et al., 1988). 

In the X-ray band, several flares have been obseved by Einstein (Vilhu \&
Linsky, 1987), 
EXOSAT (Collier Cameron et al., 1988) and GINGA (Vilhu et al., 1993). 
The flare detected in ROSAT data 
has about 350 seconds of rise time, which is much shorter than the
rise time of the flares observed by EXOSAT 
but comparable to the flares observed by GINGA. The total luminosity 
changes from about
5.8$\pm$1.6$\times10^{30}$ ergs s$^{-1}$ at the pre-flare phase to 
15.8$\pm$4.9$\times10^{30}$ ergs s$^{-1}$ at the flare
peak. The average peak luminosities of EXOSAT and GINGA flares are
(4-8)$\times10^{30}$ ergs s$^{-1}$ and (4-6)$\times10^{30}$ ergs s$^{-1}$
respectively within an energy range of 2-10 keV. While a direct comparison of 
flare luminosities is not possible since the energy range of ROSAT PSPC 
is 0.1-2.4 keV and  both EXOSAT(ME) and GINGA are sensitive to an 
energy range of about 2-10 keV, it can be inferred  that AB Dor flares 
must be soft in character, with most of the luminosity below 2 keV.
It seems that the flare develops first with  relatively high energy photons.
During the flare development the low 
energy photon flux keeps increasing while the high energy photon flux
reaches to a maximum value. The initial phase of the flare
may be due to the 
breaking of  magnetic loops which deposit their  energy to a limited
volume giving rise to the initial high energy photon flux.
As the energy deposited is dissipated in the medium an
increase in low energy photons follows. This is  characteristic of the thermal
structure of the coronal plasma observed during the flares as observed
for the solar corona (Sylwester, 1990). The higher temperature or hot 
component is present only during the flares whereas the lower temperature
or quasi-hot component might arise from the background emission within
the active region. 

A similar lead of high energy photons, followed by low energy photons may be present in
the observation of the rotational variability.
As seen from the lower panel of Figure 4, both low and high energy photons are present during the
rotational time history of AB Dor. On the active region of the surface of
the Star, several small size flares may be present giving rise to production
of low and high energy photons. But the contributions of these small size flares
or activity is integrated to a more gradual  increase
in count rate, producing the observed rotational periodicity in X-rays (Figure 3).
The GINGA light curve has been folded by the same period and it has been seen that although a broad peak is observed around phases 0.5-0.6, all the variability seen was at the 1$\sigma$ level (Vilhu et al., 1993). Figure 4 lower panel also shows that
the increase in high energy flux precedes that in the low energy flux.
Duration of the high
count rate phases indicate that for that particular observation time the
size of the active region on the star surface can be as large as $1/6$ 
of the total surface area. The rise times for the count rate increase 
in turn may be as large as 8000 seconds.  Hence, the "flares" with rise 
times $\geq$6000 seconds as observed by EXOSAT and GINGA 
in our view represents segments of the X-ray light curve corresponding to
the passage of the active region of the stellar surface of 
AB Dor through the line of sight. A true flare event is an individually
distinguished event with much smaller rise times, $\sim$100-1500 s, as
depicted in GINGA data as well as the ROSAT data analyzed here.
 ROSAT flare at phase 0.06 seems to obey the clustering trend of AB Dor
flares around phases 0.1-0.25 suggested by Vilhu et al., (1993) for
the GINGA data. The two cool
spots on the stellar surface, proposed by Innis et al. (1988), are mostly
visible around the phases 0.0 and 0.5. 
Both small and large sized X-ray flares from the corona are most likely 
linked with the same photospheric regions.
  
We note that AB Dor is similar to RS CVn binaries and other active classes
in many respects. Chromospherically active binaries with F and later spectral 
classes are known to have X-ray luminisities typically of the order of
$10^{29-31.5}$ ergs s$^{-1}$ (Pasquini et al., 1989, Dempsey et al., 1993)
which is comparable to the X-ray luminosity of AB Dor. Additionally, flare 
rise times and the peak luminosities of AB Dor and other active classes
are comparable. The rise times are 180 and 1000 seconds 
and peak flare luminosities are 9.4$\times10^{30}$ and $1.4\times10^{31}$ 
ergs s$^{-1}$ for the RS CVn binaries $\sigma^2$ CrB and HR1099 respectively. 
While the properties of AB Dor discussed above are typical 
of the active component of RS CVn binaries, it is very different from 
the typical RS CVn binaries in that it is practically a single star, being 
the active component of an extremely 
detached binary system
 ($\sim$ 250 A.U. at at distance of 
27 pc giving rise to a period of 
 $\sim$3000 years in contrast to RS CVn binaries having periods
$<$100 days.) The high level of X-ray activity of AB Dor can not be
attributed to any influence from its companion. 
Hence the X-ray activity
at the level observed in AB Dor should be intrinsic 
to the active star.
The source of the activity
is probably linked to the high equatorial rotation rate of AB Dor
and thus does not require tidal
maintenance in a binary.  
We note that in the ROSAT survey of RS CVn binaries Dempsey et al. (1993),
it was found that X-ray activity is not correlated with binary 
parameters implying that the secondary does not affect the activity level. 
\section{Acknowledgements}
We thank Hakk{\i} \"{O}gelman for his valuable suggetions, comments and 
discussion. This work is supported by The Scientific and Technical Research
Council of Turkey, under High Energy Astrophysics Unit.

\end{document}